\documentclass[letterpaper,aps,prl,twocolumn,preprintnumbers,showkeys,nofootinbib,nobibnotes]{revtex4}
\usepackage{epsfig}
\usepackage{dcolumn}% Align table columns on decimal point
\usepackage{bm}% bold math

\def\map{{{WMAP} }}
\def\cbi{{{CBI} }}
\def\acbar{{{ACBAR} }}

\newcommand{\Mpc}{\text{Mpc}}
\newcommand{\Hunit}{~\text{km}~\text{s}^{-1} \Mpc^{-1}}
\newcommand{\zre}{z_{\text{re}}}
\begin{document}  
\preprint{CITA-2003-11}

\title{Joint CMB and Weak Lensing Analysis; Physically
Motivated Constraints on Cosmological Parameters}

\author{Carlo R. Contaldi} \email{contaldi@cita.utoronto.ca}
\author{Henk Hoekstra} \altaffiliation{Department of Astronomy \&
Astrophysics, University of Toronto, 60 St. George Street, Toronto,
M5S 3H8, ON, Canada} \author{Antony Lewis} \affiliation{CITA,
University of Toronto, 60 St. George Street, Toronto, M5S 3H8, ON,
Canada}

\begin{abstract} 
We use Cosmic Microwave Background (CMB) observations together with
the Red-sequence Cluster Survey (RCS) weak lensing results to derive
constraints on a range of cosmological parameters. This particular
choice of observations is motivated by their robust physical
interpretation and complementarity. Our combined analysis, including a
weak nucleosynthesis constraint, yields accurate determinations of a
number of parameters including the amplitude of fluctuations $\sigma_8
= 0.89\pm0.05$ and matter density $\Omega_m =0.30\pm0.03$. We also
find a value for the Hubble parameter of $H_0 =70\pm3\Hunit$, in good
agreement with the Hubble Space Telescope (HST) key--project
result. We conclude that the combination of CMB and weak lensing data
provides some of the most powerful constraints available in cosmology
today.
\end{abstract} 
\date{\today} 


\maketitle 
 
%Introduction

The physics behind the anisotropies we see in the microwave background
is well studied and understood \cite{leshouches}. The evolution of the
photon distribution function in the tight coupling era and through
decoupling is well inside the linear perturbation regime and is the
reason for the CMB's unique status as a probe of cosmological
models. The {\it physical} interpretation of the angular power
spectrum of primary CMB anisotropies is unambiguous when restricted to
the inflationary paradigm and given a suitably parametrized spectrum
of initial perturbations.

The recently released \map first year results \cite{Bennett:2003bz}
have revealed the CMB angular power spectrum with unprecedented
accuracy to multipoles below $\ell=900$ \cite{Hinshaw:2003ex}.  The
results are a stunning confirmation of the acoustic oscillation
picture, with perturbations arising from an initial super-horizon
spectrum of predominantly adiabatic fluctuations, as predicted for
example by simple inflationary models.  The measurements of the first
two acoustic peaks has confirmed in precise detail earlier detections
of the peak/dip pattern on scales below the sound horizon at last
scattering \cite{toco,boom,maxima,dasi,cbi,vsa,acbar}.

On its own, the current picture of the CMB made up of the \map results
together with high resolution \cbi and ACBAR\footnote{WMAPext
combination} observations implies tight constraints on a number of
parameters \cite{Spergel:2003cb,Verde:2003ey}; the curvature in units
of critical density $\Omega_K$ and various other parameters in the
combinations determined by the physical mechanisms which give rise to
the observed CMB anisotropy.  In addition the measurement of a
cross-correlation between the polarization and temperature anisotropy
\cite{Kogut:2003et,Page:2003fa} is the first significant detection of
reionization in the CMB, which gives a constraint on the optical depth
to the last scattering surface.

Although the CMB data alone provide tight constraints on some
parameter combinations, other combinations are very poorly constrained
due to partial degeneracies. The addition of other data such as
measurements of the matter power spectrum $P(k)$ is essential to break
these degeneracies and tightly constrain the parameters of most
interest individually. One way to infer the matter power spectrum is
to rely on visible tracers of the (dark) matter distribution such as
galaxy redshift surveys or observations of the Lyman-$\alpha$ forest.
The Lyman--$\alpha$ forest gives a way to measure the linear power
spectrum of neutral gas at redshifts higher than those probed by
galaxy surveys.

The combination of CMB, 2dFGRS \cite{Colless:2001gk}, and
Lyman--$\alpha$ forest data \cite{croft2002} yields tight constraints
on the density of dark matter and vacuum energy, and also reveal an
indication of a running of the scalar spectral index characterized by
the parameter $dn_s/d\ln k$ \cite{Spergel:2003cb}. However inferring
the matter power spectrum using these techniques involves a heuristic
treatment of the relation between the tracers and the dark matter
usually referred to as `biasing' \cite{kaiser}. As we enter the much
heralded era of precision observations, such heuristic treatments
might limit the accuracy with which parameters can be determined. A
direct measurement of the power spectrum would not suffer from such
limitations.

In terms of physical interpretation, measurements of the lensing
signal induced by the LSS (cosmic shear) hold a unique position in the
growing set of observational tools available to cosmologists; it is a
{\it direct} probe of the projected matter power spectrum over a
redshift range determined by the lensed sources and over scales
ranging from the linear to non--linear regime. The intervening LSS
induces a small, coherent correlation in the shapes of the background
galaxies which nowadays can be measured accurately
\cite{Bacon:2000sy,Hoekstra:2002cj,Kaiser:2000if,Ludo2000,vanWaerbeke:2002pw}. The
use of weak lensing data is not without challenges: the small signal
requires large survey areas and a careful removal of the observational
distortions. However separation of the shear signal into gradient
(``E-Type'') and curl (``B-Type'') components provides a control on
systematics including the presence of intrinsic alignments of nearby
galaxies or 
%any 
systematically induced distortions in the image. The
RCS 53 sq. deg. results used in this work \cite{Hoekstra:2002xs} have
a low B-Type component on large scales together with a well determined
redshift distribution of background sources.

In this {\it letter} we present results from cosmological parameter
fits using only CMB and weak lensing data. The motivation for this
approach is to provide constraints on parameters using only
observables with robust physical interpretations.

%method

To evaluate the posterior distribution of the parameters of interest
from the data we use an extension of the publicly
available Markov Chain Monte Carlo package \textsc{cosmomc}\footnote{http://cosmologist.info/cosmomc/},
as described in \cite{Lewis:2002ah}. We calculate the likelihood
of each cosmological model with respect to a combination of CMB and
RCS data. The CMB data consists of \map data below $\ell=900$ and
\cbi, \acbar, and VSA band powers above $\ell=800$
where the \map data is noise dominated and hence the band powers are
essentially independent.
To compare each angular power spectrym to the WMAP data we use the
likelihood calculation routine made available by the WMAP
team\footnote{http://lambda.gsfc.nasa.gov/} \cite{Verde:2003ey}.

For each model we also calculate the mass aperture variance $\langle
M_{ap}^2(\theta)\rangle$ \cite{Schneider:1997ge} at each aperture
$\theta$ sampled by the RCS results \cite{Hoekstra:2002xs}. The mass
aperture variance is a narrow filter of the convergence power spectrum
$P_{\kappa}(\ell)$ defined as
\begin{equation}\label{eq:p_kappa}
        P_{\kappa}(\ell) = \frac{9}{4}\left(\frac{H_0}{c}\right)^4\Omega_m^2\int^{\chi_H}_{0} d\chi \frac{g^2(\chi)}{a^2(\chi)}P_{3D}\left(\frac{\ell}{f_{K}(\chi)};\chi\right),
\end{equation}
where $\chi$ is the radial coordinate and $f_{K}(\chi)$ is the
comoving angular diameter distance to $\chi$. $P_{3D}(k,\chi(z))$is
the 3D power spectrum of matter fluctuations. For each model we use
the matter power spectrum calculated by \textsc{camb}
\cite{Lewis:1999bs} at $z=0$ and rescale to $z>0$ using the solution
for growth of linear perturbations. To include the non--linear
contribution to the power spectrum at each redshift we use the
\textsc{halofit} procedure\cite{Smith:2002dz}. The procedure has been
calibrated using numerical simulations of structure formation and is
significantly more accurate than the previous procedure by Peacock \&
Dodds \cite{pd96}. In particular it reproduces accurately, with {\it
rms} errors of a few percent, the full non--linear spectrum in
standard $\Lambda$CDM models down to scales $k\sim 10 h$Mpc$^{-1}$.
%,with larger
%errors for open CDM models which we do not consider in this work. 
The accuracy of the \textsc{halofit} procedure is adequate for current
weak lensing data although future surveys will require more accurate
estimates of the full, non--linear power spectrum. This will most
probably require the use of large numbers of numerical simulations to
calibrate directly the non--linear evolution in the full parameter
space.

The function $g(\chi) = \int^{\chi_H}_{\chi} d\chi' p(\chi')  f_K(\chi'-\chi)/f_K(\chi')$ is
the source--averaged distance ratio
%\begin{equation}
%        g(\chi) = \int^{\chi_H}_{\chi} d\chi' p(\chi') \frac{f_K(\chi'-\chi)}{f_K(\chi')},
%\end{equation}
where $p(\chi(z))$ describes the redshift distribution of sources in
the shear survey which is approximated by the function $p(z) \sim (z/z_s)^{\alpha}\exp\left[-(z/z_s)^{\beta}\right]$.
The values $\alpha=4.7$, $\beta=1.7$, and $z_s=0.302$
give the best fit to the observed redshift distribution. To allow for
the uncertainty in the mean redshift of the distribution we
marginalize over the range of values $z_s \in [0.274,0.337]$ for each
likelihood evaluation. This corresponds to the $\pm 3\sigma$ range
indicated by the $\chi^2$ of the fit to the photometric redshift
distribution. The mean redshift for this choice of parameters is
$\langle z \rangle = 0.54-0.66$. We assume a Gaussian prior for
$z_s$ in this range.
%A possibly comment on sensitivity to redhift distribution and priors
%on z_s, Jarvis result, etc..

For each model sampled by the Monte Carlo chain we calculate the
log likelihood with respect to the RCS data
\begin{equation}
        \ln L = -\frac{1}{2}\left( \widetilde{\langle M^2_{ap}\rangle}_i-\langle M^2_{ap}\rangle_i\right)C^{-1}_{ij}\left( \widetilde{\langle M^2_{ap}\rangle}_j-\langle M^2_{ap}\rangle_j\right),
\end{equation}
where $\widetilde{\langle M^2_{ap}\rangle}_i$ is the observed mass
aperture variance at an aperture $\theta_i$ and $C_{ij}$ is the
covariance matrix of the data \cite{Hoekstra:2002xs}. This result is
added to the log likelihoods from the CMB fit for the same model to
obtain the full likelihood with respect to both CMB and RCS data.

\begin{figure}[ht]
\centerline{\psfig{file=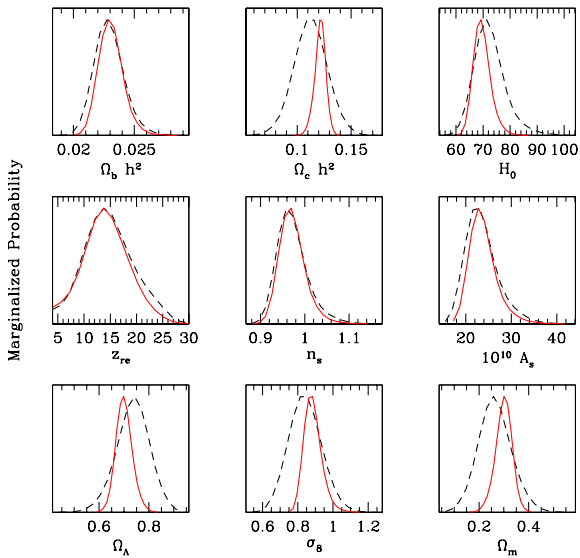,width=8 cm,angle=0}} 
\caption{\label{fig:fig1} One dimensional, marginalized probability
distributions for a selection of parameters. The dashed (black) curves
are for CMB only. The solid (red) curves include the RCS data. We see
the weak lensing data, while being consistent with the CMB only
results, narrow down a number of distributions considerably. In
particular the combination of data provides a tight independent
constraint on $H_0$. The matter density and fluctuation amplitude are
also much better constrained with the combination of CMB and weak
lensing than with just CMB.}
\end{figure}

We sample the probabilities with respect to six basic cosmological
parameters: the physical densities of baryons $\Omega_bh^2$, and cold
dark matter $\Omega_c h^2$, the Hubble parameter $H_0 \equiv 100h
\Hunit$, a reionization redshift parameter $\zre$, and a constant
spectral index $n_s$ and amplitude $A_s$ of the initial scalar
curvature perturbations.  We assume the universe is spatially flat,
with purely adiabatic perturbations evolving according to General
Relativity.
%We assume the universe's background
%geometry is flat with $\Omega_{\rm tot} = 1.0$, so 
The density of a
cosmological constant type component $\Omega_{\Lambda}$ 
%then 
follows from $\Omega_{\Lambda} = 1.0 - \Omega_m$.
%, with purely adiabatic
%perturbations evolving according to General Relativity.  
We generated sixteen converged Monte Carlo chains using the CMB data
only, removed burn in and thinned to obtain fairly independent
samples.  The matter power spectrum and RCS likelihood was then
computed for each sample, and importance sampling used to adjust the
chain weights accordingly (see~\cite{Lewis:2002ah}). The resultant set
of weighted samples for the full posterior distribution from the CMB
and RCS data were then used to compute our results.  The only external
prior assumed is a conservative big bang nucleosynthesis (BBN)
Gaussian prior of the form $\Omega_bh^2=0.022\pm0.002$ (1$\sigma$)
\cite{Tytler:2000qf}. We include this prior to partially break the
remaining $n_s$--$\Omega_bh^2$--$\tau$--$A_s$ degeneracy in the CMB
data. The action of this is similar to the $\tau<0.3$ prior adopted in
the WMAP analysis \cite{Spergel:2003cb,Verde:2003ey}.

From the set of samples it is simple to also compute the posterior
distribution of other derivable quantities such as the {\it rms}
amplitude of matter fluctuations on $8h^{-1}$Mpc scales assuming
linear evolution, $\sigma_8$, the total matter density, $\Omega_m$,
the optical depth to last scattering, $\tau$, and the age of the
universe. In this letter we do not consider tensor perturbations,
dynamical dark energy candidates, or a running spectral index.  We
will explore the constraints on these generalized models from CMB and
weak lensing data future work.

%Results 

The set of samples from the full six dimensional parameter space can
be used to evaluate marginalized parameter distributions by evaluating
the weighted number density of samples ignoring the values of the
parameters marginalized over.  In Fig.~\ref{fig:fig1} we show the one
dimensional marginalized distributions for a number of parameters.
Each panel compares the distribution obtained using CMB data with that
obtained using CMB and RCS data together; both also include the weak
BBN prior discussed above.  The effect of adding the weak lensing
results is clearly seen in a number of parameters.

In Table~\ref{tab:table1} we summarize the marginalized constraints
for a number of fundamental and derived parameters. We show the
results obtained with and without inclusion of the RCS data.  The
addition of RCS data reduces the errors on $\sigma_8$, $\Omega_m$,
$H_0$, $\Omega_{\Lambda}$, and $\Omega_ch^2$. We also show constraints
on the `classical' combinations probed by LSS data, namely, the
constrained direction $\sigma_8\Omega_m^{0.5}$ and the shape parameter
$\Gamma\approx\Omega_mh$.

\begin{table}
\caption{\label{tab:table1} Marginalized constraints for a selection
of parameters. The left column uses only CMB data, the right column is
for CMB and RCS data. The only external prior included for both
results is a Gaussian BBN prior of $0.022\pm0.002$.
All errors are $68\%$ confidence intervals.}
\begin{ruledtabular}
\begin{tabular}{ccc}
($\Omega_{tot}=1$)&BBN+CMB\footnote{WMAP($\ell<900$) + CBI,ACBAR,VSA($\ell>800$)}&BBN+CMB$^a$+RCS\\
\hline
$\Omega_bh^2$           & $0.023\pm0.001$       &  $0.023\pm0.001$ \\
$\Omega_c  h^2$        & $0.112\pm0.016  $       &  $0.121\pm0.005 $ \\
$h$                      & $0.73\pm0.06   $       &  $0.70\pm0.03   $ \\
$\zre$                 & $15\pm5   $       &  $15\pm4   $ \\
$n_s$                   & $0.97\pm0.03  $       &  $0.97\pm0.03   $ \\
$10^{10} A_s$           & $24\pm4   $       &  $25\pm3   $ \\
\hline
$\Omega_{\Lambda}$      & $0.74\pm0.07  $       &  $0.70\pm0.03   $ \\
$\Omega_m$              & $0.26\pm0.07  $       &  $0.30\pm0.03  $ \\
$T_0$(Gyrs)             & $13.6\pm0.3   $       &  $13.6\pm0.2   $ \\
$\sigma_8$              & $0.84\pm0.09  $       &  $0.89\pm0.05  $ \\
$\sigma_8e^{-\tau}$              & $0.73\pm 0.08$        &  $0.78\pm 0.02 $ \\
$\sigma_8\Omega_m^{0.5}$ & $0.43\pm0.09$          &  $0.48\pm0.02$ \\
$\Omega_mh$             & $0.19\pm 0.03$                 &
$0.21\pm0.01$ \\
$\Omega_m h^{2.3} (\sigma_8 e^{-\tau})^{-0.9}$ & $ 0.163\pm0.003$ & $
0.162\pm 0.002$ \\
\end{tabular}
\end{ruledtabular}
\end{table}

\begin{figure}[t]
\centerline{\psfig{file=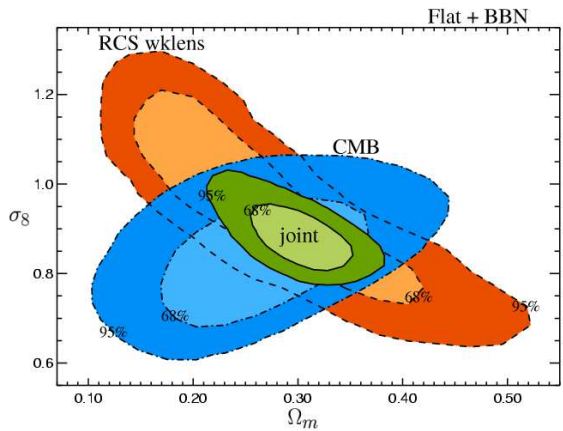,width=8 cm,angle=0}} 
\caption{\label{fig:s8om} The two dimensional, marginalized
likelihoods for the $(\Omega_m,\sigma_8)$ plane. The overlaid, filled
contours show the 68\% and 95\% integration levels for the
distributions. Bottom -- RCS only, Middle -- CMB only, Top --
CMB+RCS.} 
\end{figure}

It is instructive to look at the marginalized, two dimensional
likelihood in the ($\Omega_m,\sigma_8$) plane to understand how
drastic improvements in the determination of the two parameters are
obtained (Fig.~\ref{fig:s8om}). The RCS data alone is near degenerate
in a particular direction while CMB data alone provides broad
constraints in a {\it quasi}--orthogonal direction to RCS. The
combination of the two data sets give a much tighter confidence region.
The region of intersection in six dimensions has slightly above
average CMB likelihood, as is readily assessed using the importance
weighted samples, so the data sets are highly consistent even in the
full parameter space. The spread of the CMB posterior in the
direction of the RCS degeneracy is largely due to the uncertainty
remaining in the optical depth, as is clear for the tight constraint for
$\sigma_8 e^{-\tau}$ given in Table~\ref{tab:table1}. 
 
%Discussion
Overall our results are consistent with similar constraints from a
combination of CMB, 2dFGRS, and Lyman--$\alpha$ data
\cite{Bennett:2003bz,Spergel:2003cb} with similar or smaller
errors.
The values obtained for $\sigma_8$ using the WMAP data are somewhat
higher than those obtained previously from CMB data due to the new evidence for a
significant optical depth and a slightly higher
anisotropy amplitude than previous observations indicated~\cite{Hinshaw:2003ex}. 
This is still lower, although not inconsistent, with estimates of
$\sigma_8$ from a possible Sunyaev-Zeldovich Effect (SZE) contribution
to the CMB power spectrum at high-$\ell$. The latest estimates using
the CBI deep--field results \cite{Mason:2002tm,Bond:2002tp} and ACBAR and BIMA
\cite{Dawson:2002dg} data suggests a value of
$\sigma_8^{SZ}=0.98^{+0.12}_{-0.21}$ \cite{Goldstein:2002gf} with
large errors due mainly to the non-Gaussian nature of the
SZE. Increasingly accurate measurements of the CMB power spectrum at
high-$\ell$ will reduce these errors drastically and comparing the two
independent determinations of $\sigma_8$ will be useful in increasing
our understanding of the cluster properties that determine the SZE.

Our result for the Hubble parameter is consistent within 1$\sigma$
with the HST key--project result \cite{Freedman:2000cf} but has
smaller errors. Similarly, the value for the matter density $\Omega_m =
0.30\pm0.03$ is consistent with other determinations. The addition of
RCS data leave estimates of the scalar spectral tilt $n_s$ essentially
unaffected. This is due to the small range of scales probed by the RCS
weak lensing results. Future surveys will most certainly have much
more leverage on $n_s$ as they will probe a range in scales an order
of magnitude larger.

We have shown how CMB and weak lensing results can be combined to
obtain constraints on cosmological parameters that depend on
observations that have simple physical interpretations. Although only
first generation weak lensing data are currently available our
approach yields results with errors comparable to or even smaller than
those obtained using CMB in combination with other types of
surveys. These results are encouraging for the use of next generation
weak lensing surveys in deriving robust parameter fits. In particular
the Canada--France--Hawaii--Telescope (CFHT) Legacy Survey $\sim 170
$ sq. deg. cosmic shear project will be a major step forward in the
field of weak lensing.

The increasing
accuracy in the determination of the source redshift distribution in
future surveys will also help in reducing uncertainties and
systematics tied to any intrinsic alignment in the ellipticity of
nearby sources. It will also introduce the possibility of resolving
separate redshift contributions to the convergence power spectrum
(Eq.~\ref{eq:p_kappa}) thus enhancing the parameter fitting ability of
the observations.

We conclude that the combined CMB, weak lensing approach to parameter
determination already constitutes a competitive alternative to other
combinations and holds much promise for future investigations. 

%Acknowledgments
 
It is a pleasure to thank Dick Bond, Ue-Li Pen, and Dmitry Pogosyan
for useful discussions. We acknowledge the RCS team for use of their
data. Research at CITA is supported by NSERC and the Canadian
Institute for Advanced Research. The computational facilities at CITA
are funded by the Canadian Fund for Innovation. CRC acknowledges the
Kavli Institute for Theoretical Physics (National Science Foundation
Grant No. PHY99-07949) where part of this work was carried out.

\end{document}